\documentclass[iop]{emulateapj-rtx4}
\usepackage{natbib}
\usepackage{graphicx}                    
\usepackage{amssymb,amsmath}
\usepackage[usenames,dvipsnames]{color}
\usepackage{rotating}

\newcommand{\BE}{\begin{equation}}
\newcommand{\EE}{\end{equation}}
\newcommand{\BA}{\begin{eqnarray}}
\newcommand{\EA}{\end{eqnarray}}
 \newcommand{\fig}[1]{Figure~\ref{#1}}

 \newcommand{\sect}[1]{Section~\ref{#1}}

\newcommand{\angstrom}{\mbox{\normalfont\AA}}

\shorttitle{Active region FIP bias evolution}
\shortauthors{Baker et al.}

\begin{document}

\title{FIP bias evolution in a decaying active region} 

\author{D. Baker\altaffilmark{1}, D.~H. Brooks\altaffilmark{2}, P.~ D\'emoulin\altaffilmark{3}, S.~L.~Yardley$^{1}$, L. van Driel-Gesztelyi\altaffilmark{1,3,4},   D.~M. Long\altaffilmark{1}, L.~M. Green \altaffilmark{1}}
\altaffiltext{1}{University College London, Mullard Space Science Laboratory, Holmbury St Mary, Dorking, Surrey, RH5 6NT, UK}
\altaffiltext{2}{College of Science, George Mason University, 4400 University Drive, Fairfax, VA 22030, U.S.A.}
\altaffiltext{3}{Observatoire de Paris, LESIA, CNRS, UPMC Univ. Paris 06, Univ. Paris-Diderot, Meudon, France}
\altaffiltext{4}{Konkoly Observatory, Hungarian Academy of Sciences, Budapest, Hungary}

\begin{abstract}
Solar coronal plasma composition is typically characterized by first ionization potential (FIP) bias.  Using spectra obtained by \emph{Hinode's} EUV Imaging Spectrometer (EIS) instrument, we present a series of large-scale, spatially resolved composition maps of active region (AR)~11389.  The composition maps show how FIP bias evolves within the decaying AR from 2012 January 4--6.  Globally, FIP bias decreases throughout the AR.  We analyzed areas of significant plasma composition changes within the decaying AR and found that small-scale evolution in the photospheric magnetic field is closely linked to the FIP bias evolution observed in the corona.  During the AR's decay phase, small bipoles emerging within supergranular cells reconnect with the pre-existing AR field, creating a pathway along which photospheric and coronal plasmas can mix.  The mixing time scales are shorter than those of plasma enrichment processes.  Eruptive activity also results in shifting the FIP bias closer to photospheric in the affected areas.  Finally, the FIP bias still remains dominantly coronal only in a part of the AR's high-flux density core.  We conclude that in the decay phase of an AR's lifetime, the FIP bias is becoming increasingly modulated by episodes of small-scale flux emergence, i.e. decreasing the AR's overall FIP bias.  Our results show that magnetic field evolution plays an important role in compositional changes during AR development, revealing a more complex relationship
than expected from previous well-known Skylab results showing that FIP bias increases
almost linearly with age in young ARs (Widing $\&$ Feldman, 2001, \emph{ApJ}, 555, 426).

\end{abstract}
\keywords{Sun:abundances---Sun:active region---solar wind }

\section{Introduction}
\label{intro}

Knowledge of plasma composition is key to understanding the physical processes associated with how solar plasma is transported, heated, and accelerated in the solar atmosphere.  In order to characterize composition variations, the first ionization potential (FIP) bias (FIP$_{Bias}$) is used to define the ratio of the elemental abundance in the solar atmosphere (A$_{SA}$) to the elemental abundance in the photosphere (A$_{Ph}$) such that FIP$_{Bias}$ = A$_{SA}$/A$_{Ph}$.  Composition is a critical property that distinguishes the slow from fast solar wind \citep[SW;][]{geiss95,vonsteiger98,antiochos11}.  The fast, steady wind has stable photospheric composition i.e. FIP bias $\sim$1 \citep{vonsteiger98,zurbuchen98,vonsteiger01,zurbuchen02}, and the slow, non-steady wind has variable coronal composition characteristic of AR plasma or high FIP bias \textgreater 3--4 \citep[e.g.][]{zurbuchen06}. Further results of abundance variations in discrete coronal structures were summarized in Table~1 of \cite{baker13} and in review papers, such as \cite{feldman03}.

Plasma confined in closed-loop structures at the time of emergence from beneath the photosphere has been determined to have photospheric composition \citep{sheeley95,sheeley96,widing97,widing01}. However, once emerged, the plasma composition appears to be modified within days.  \cite{mckenzie92} and \cite{saba93} analyzed more established ARs in soft X-rays and found variations in Fe/Ne (low/high FIP elements) abundance ratios ranging from 2--7.  In EUV, \cite{widing95} derived FIP bias values of 4.8, 5.4 and 5.9 for three ARs.  

FIP bias levels in some older ARs have been measured from 8--16  \citep{young97,dwivedi99,widing01,feldman03}.  \cite{widing01} presented a systematic study of compositional evolution in ARs using the \emph{Skylab} spectroheliograph observations.  Four ARs were identified from 1973/4, during the minimum of Solar Cycle 20, which emerged on the visible side of the Sun and could be followed for 3--7 days.  Using Mg VI/Ne VI intensity ratios, \cite{widing01} demonstrated that plasma trapped in newly emerged loops has photospheric composition and from thereon the AR plasma becomes enriched at an almost constant rate per day so that after 2--3 days coronal and slow SW abundances of 3--4 were reached.  After 4--5 days, FIP bias had risen to 8--9.  Based on their study, \cite{widing01} concluded that there is a close relationship between the duration of plasma confinement and the magnitude of the FIP bias.  Surprisingly, few studies have been conducted on the evolution of FIP bias of ARs after emergence beyond that of \cite{widing01}.

In this paper, we examine the evolution of FIP bias in a decaying AR that is older than those studied in \cite{widing01}.  AR~11389 was well observed from 2012 January 4 to 6 by several instruments (\sect{obs}).  In \sect{Decay_phase} we first describe the photospheric and coronal observations of AR~11389 and its nearby coronal hole (CH).  This is followed in \sect{results} by a detailed investigation of the large-scale changes in FIP bias within the supergranular cell forming part of the AR's negative polarity as well as a consideration of the small-scale, localized  changes in FIP bias in and around the AR.  We finish in \sect{discussion} with a discussion of the implications of our findings for the long-held view that AR FIP bias increases linearly with time before concluding in \sect{Conclusion}.

\section{Observations}
\label{obs}
\begin{figure*}[p]   
\epsscale{0.85}
\plotone{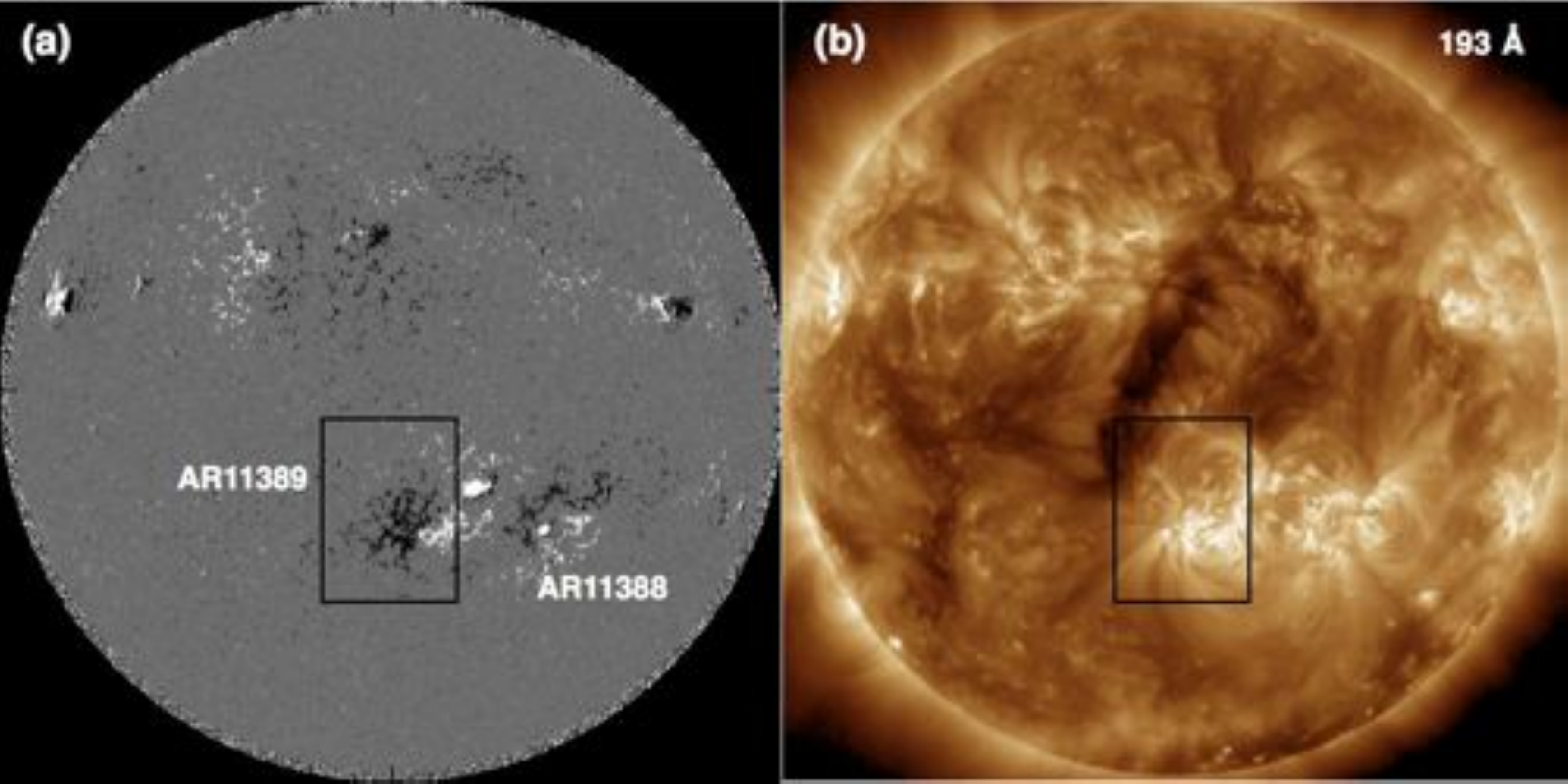}
\plotone{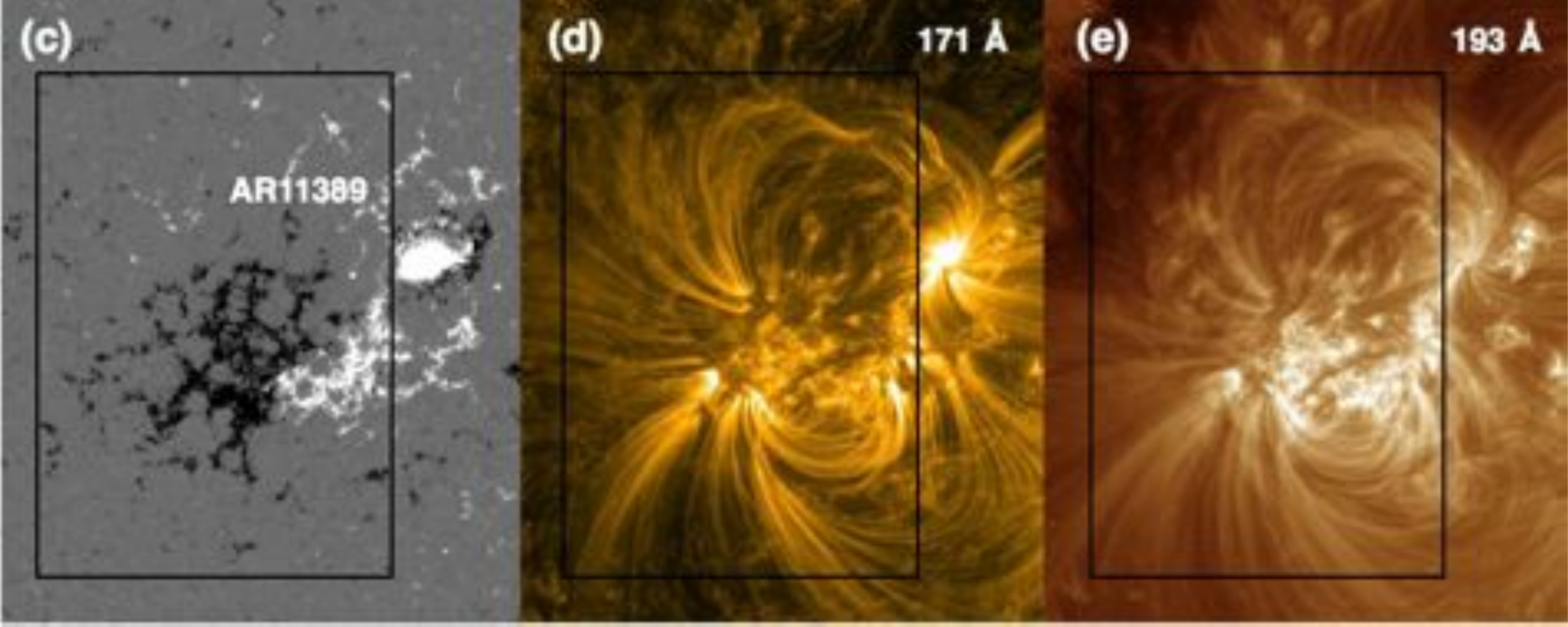}
\caption{Context images of AR-CH complex on 2012 January 4 at 08:00 UT.   (a) Full disk SDO/HMI line-of-sight magnetogram saturated at $\pm$ 200 G and (b) SDO/AIA 193 \AA\ image.  (c) Zoomed SDO/HMI, (d) SDO/AIA 171 \AA\ and (e) 193 \AA\ high-resolution images \citep[processed using the Multiscale Gaussian Normalization (MGN) technique of][]{morgan14}.  Black box indicates the  \emph{Hinode}/EIS field of view which data are shown in \fig{all_eis_hmi}.
 \label{context}}.
\end{figure*}

The observations presented here were obtained during 2012 January 4--6 using the \emph{Hinode} Extreme-ultraviolet Imaging Spectrograph \citep[EIS:][]{culhane07},  the Solar Dynamics Observatory (SDO)/Atmospheric Imaging Assembly \citep[AIA:][]{lemen12} and SDO/Helioseismic and Magnetic Imager \citep[HMI:][]{schou12,scherrer12} instruments.  \fig{context}a,b show full-disk  HMI and SDO/AIA 193 \AA\ context images on the 4th at 08:00 UT.  The CH is located close to solar central meridian (CM) with AR~11389 on its southwestern boundary at approximately 350$\arcsec$ south of the solar equator.  AR 11388 is west of the AR-CH complex.  Zoomed SDO/HMI, SDO/AIA 171 \AA\ and 193 \AA\ images of AR~11389 overplotted with the EIS field of view (FOV) are shown at the bottom of \fig{context}.

\emph{Hinode}/EIS observed AR~11389 using the slit scanning mode with the 2$\arcsec$ slit and 2$\arcsec$ scan step size for 180 pointing positions to build up a raster with a large FOV of 360$\arcsec$$\times$512$\arcsec$ on January 4 and 6.  A smaller sparse raster FOV of 195$\arcsec$$\times$280$\arcsec$ was constructed using the same slit but with a scan step size of 3$\arcsec$ for 65 pointing positions on January 5.  

\begin{figure*}   
\epsscale{0.9}
\plotone{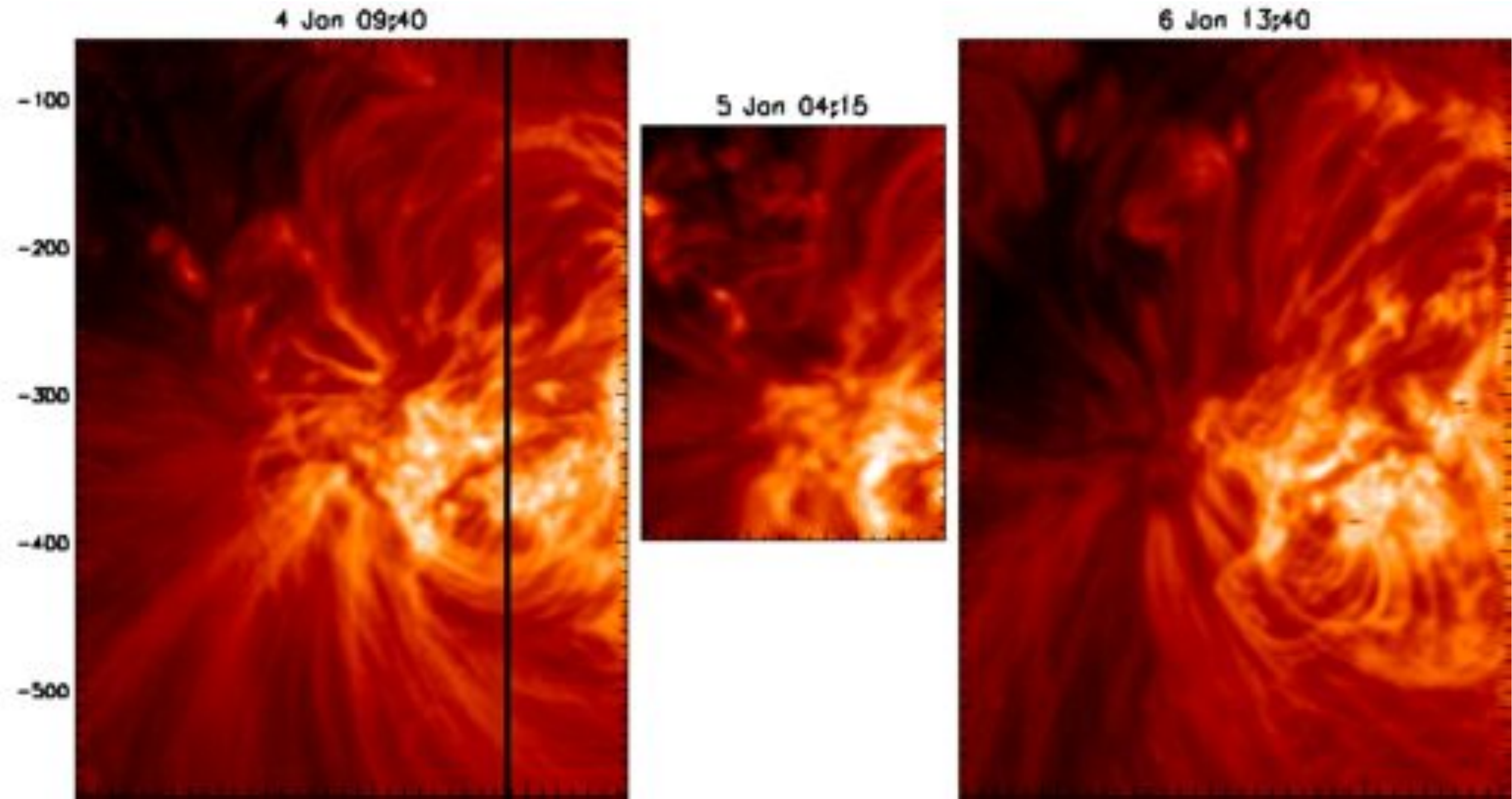}
\plotone{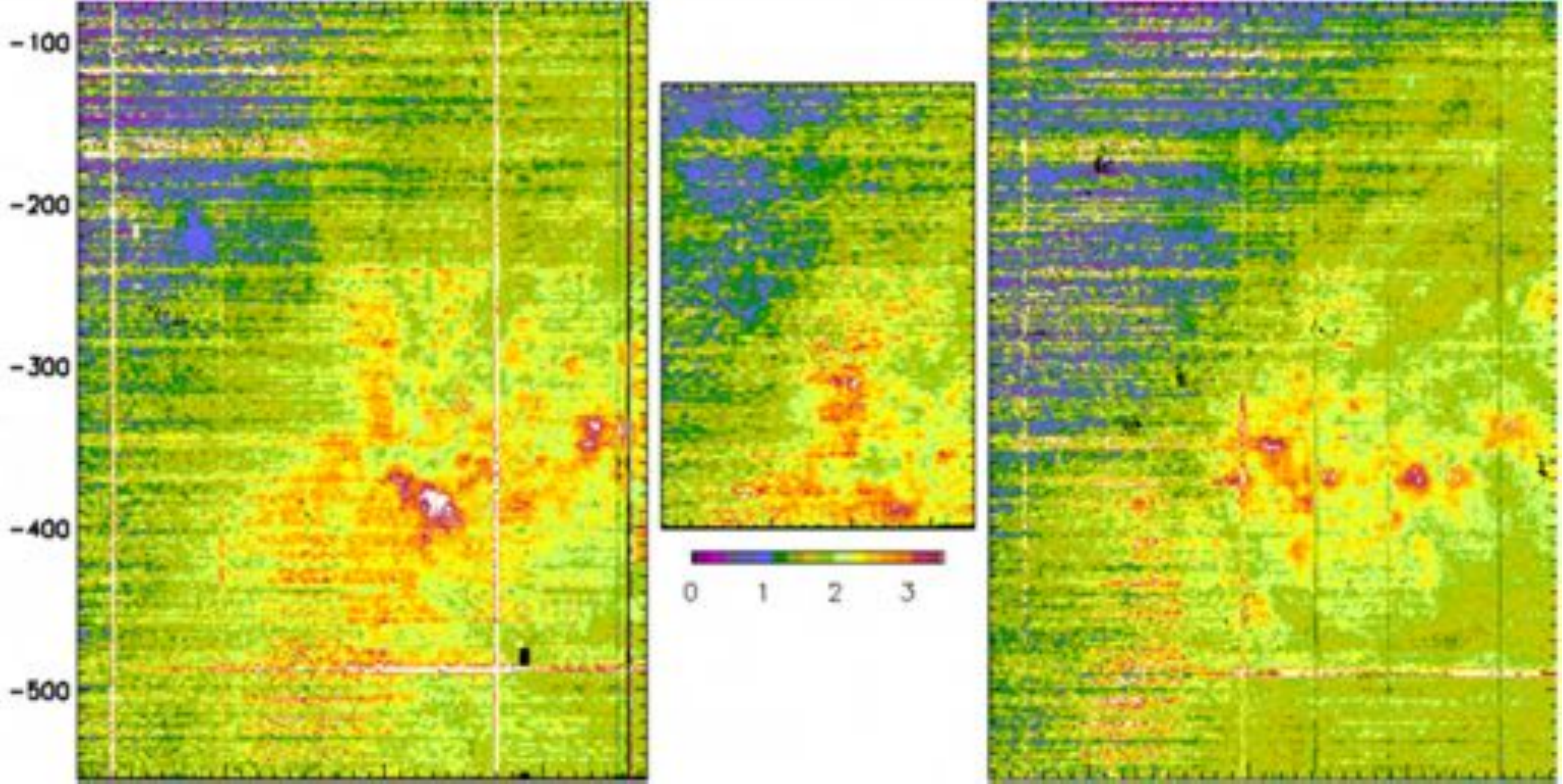}
\plotone{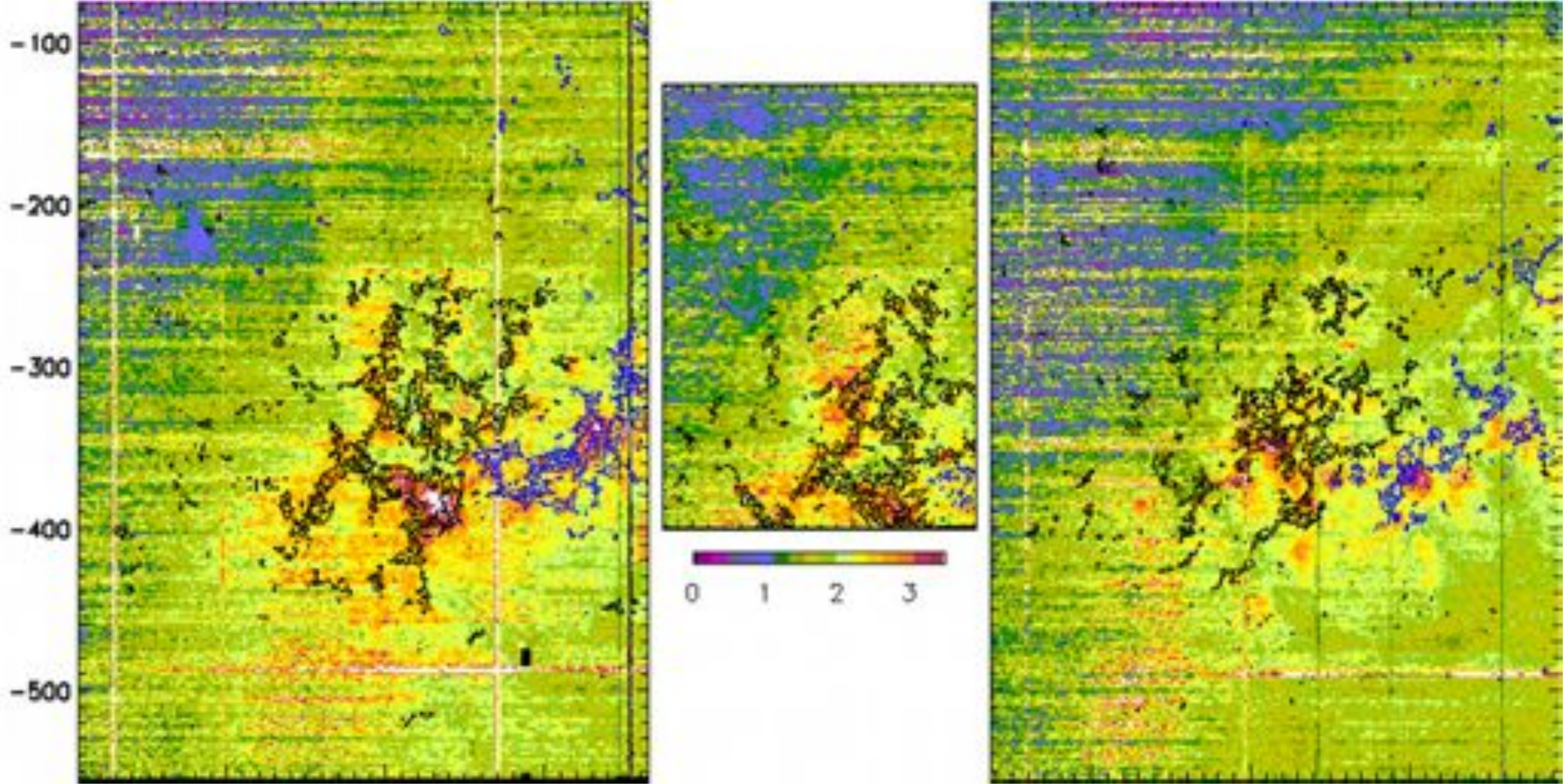}
\caption{(Top to bottom) \emph{Hinode}/EIS Fe {\sc xii} 195.12 {\AA} high-resolution intensity \citep[processed using MGN technique of][]{morgan14}, S {\sc x} 264.223 {\AA} -- Si {\sc x} 258.375 {\AA} composition  maps, without/with SDO/HMI magnetogram contours of $\pm$200 G (blue/black contours) for (left to right) 2012 January 4 at 09:40 UT, 2012 January 5 at 04:15 UT, 2012 January 6 at 13:40 UT.  }\label{all_eis_hmi}
\end{figure*}

\fig{all_eis_hmi} shows Fe {\sc xii} 195.12 \AA\ intensity maps and S {\sc x} 264.223 \AA\ -- Si {\sc x} 258.375 \AA\ composition (or FIP bias) maps, with and without SDO/HMI magnetogram contours of $\pm$200 G overplotted.  Standard SolarSoft EIS procedures were used to correct the raw data for dark current, cosmic rays, hot, warm, and dusty pixels and to remove instrumental effects of orbital variation, CCD detector offset, and slit tilt.  Single Gaussian functions were used to fit the unblended calibrated spectra
for
Fe {\sc viii} 185.213 $\angstrom$, 
Fe {\sc ix} 188.497 $\angstrom$,
Fe {\sc x} 184.536$ \angstrom$,
Fe {\sc xiii} 202.044 $\angstrom$,
Fe {\sc xiv} 264.787 $\angstrom$,
Fe {\sc xv} 284.160 $\angstrom$,
Fe {\sc xvi} 262.984 $\angstrom$,
S {\sc x} 264.233 $\angstrom$,
and Si {\sc x} 258.375 $\angstrom$,
while double Gaussian functions were used to separate the blended
Fe {\sc xi} 188.216 $\angstrom$,
Fe {\sc xii} 195.119 $\angstrom$,
and Fe {\sc xiii} 203.826 $\angstrom$
lines. All these lines were used in the construction of the composition
maps.  A discussion of the method used to make these maps may be found in \cite{brooks11} and \cite{baker13}.   One caveat with this study is that the composition maps reduce the
data in an identical fashion for all features in the AR. We have not,
for example, isolated the loops from the background which must be kept in mind when analyzing changes in composition maps from January 4--6.

\section{Decay phase of AR~11389}
\label{Decay_phase}
AR~11389 appeared at the east limb on 2011 December 28.  The AR was comprised of two positive polarity spots and dispersed following negative field for the first few days of its solar disk transit.  By the time it approached the CM, the southern spot had decayed.  The magnetic field configuration at CM is displayed in \fig{context}c.  Five M-class GOES X-ray flares were attributable to the AR prior to January 2.  Activity then decreased to intermittent C-class flares.  

The precise age of AR~11389 can not be determined from direct observation of its emergence.  However, based on a number of factors, we can reasonably state that the AR is significantly older than those selected by \cite{widing01} in their analysis of FIP bias evolution in young ARs.  First, AR~11389 is visible in STEREO B at the east limb on 2011 December 20, 15 days prior to the first \emph{Hinode}/EIS observation on January 4.  It is visible in STEREO A approximately one week earlier.  Second, during its on-disk transit, the level of flaring in the AR had reduced to very low levels, suggesting its age was beyond that of the flux emergence phase.  Typically, flare activity peaks when an AR's sunspot area is at its maximum development just before the decay phase \citep{waldmeier55,choudhary13,lvdg14}.  Third,
the lifetimes of ARs are approximately proportional to their peak magnetic flux \citep{schrijver00}.  AR~11389 is a large AR with total negative magnetic flux measured to be $\sim$1.3$\times$10$^{22}$ Mx at CM.  Its lifetime is likely to be weeks--months rather than days--weeks \citep{lvdg14}.  All of these points imply that AR~11389 is expected to be older than two weeks but less than one month as it was not present during the previous solar rotation.
\section{FIP Bias Evolution in AR~11389}
\label{results}
\begin{figure}   
\epsscale{1.0}
\plotone{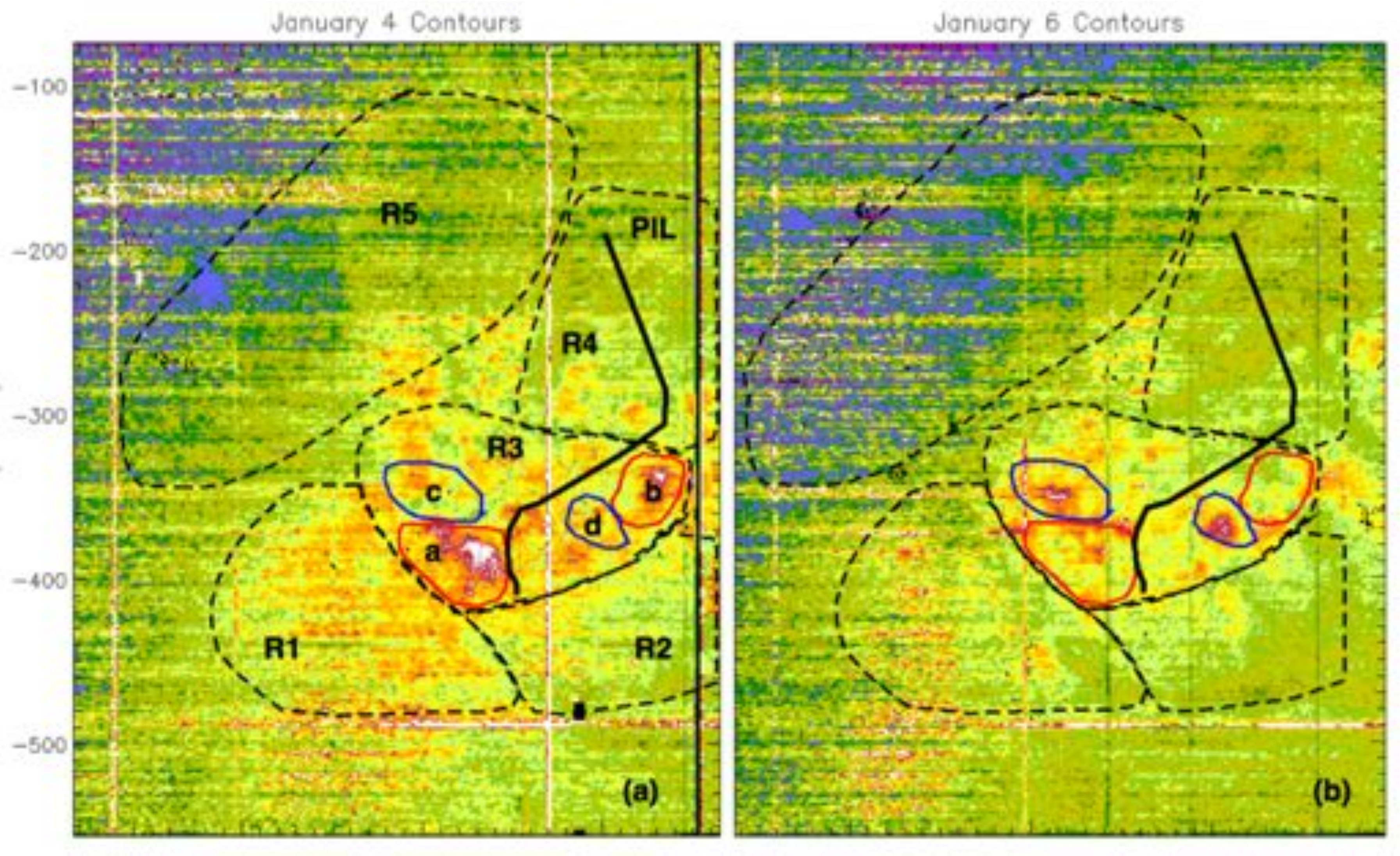}
\caption{Composition maps overplotted with contours designating major regions R1--R5, subregions 3a--3d, and the main polarity inversion line (PIL; solid black lines) on 2012 Jan 4 at 09:40 UT and 2012 Jan 6 at 13:40 UT. \label{contours}}
\end{figure}

We give a detailed account of FIP bias evolution within AR~11389 in the following subsections.  Based on the composition maps, the AR has been divided into regions R1--R5, indicated by the dashed contours in Figure \ref{contours}.  The positions  of the contours within the EIS FOV change slightly from January 4--6 as the AR evolves so the polarity inversion line (PIL) was used as a fixed reference for the placement of the contours on the 6th.  The underlying magnetic field and SDO/AIA multi-wavelength images are used to understand the changes in plasma composition observed in the upper atmosphere from the EIS raster at 09:40 UT on January 4 to 13:40 UT on January 6.  Frequent references are made to the two movies included as supplemental on-line material:  FIP$\_$EIS$\_$AIA.mov and SDO$\_$multi$\_$panels.mov. 

\subsection{Composition Evolution: Supergranular Cell}
\label{results_Supergranular}

The most striking change observed in the composition maps of AR~11389 over two days is the decrease in FIP bias in R1 within the negative field of the AR (\fig{contours}).  The probability distribution functions (PDFs) of FIP bias pixel$^{-1}$ within R1 for January 4th (black PDF) and 6th (red PDF) are shown in Figure \ref{pdfs}, as are the PDFs for all regions.  In R1, the PDF of FIP bias pixel$^{-1}$ has shifted to lower levels, \textit{i.e.}~towards photospheric composition, from January 4--6 and the mean value has decreased from 2.30 to 2.04.  

As is evident in the SDO/HMI magnetograms in the accompanying SDO$\_$multi$\_$panels.mov, evolution of the magnetic field within R1 is driven largely by small scale flux emergence causing fragmentation and merging around the supergranular cell's periphery.  We identified the primary sites of flux emergence events for the two-day period using an automated magnetic flux detection algorithm.  Contours were fitted around the emerging positive flux fragments employing a clumping identification method with a contour level of 30 G (3$\sigma$ accuracy) \citep{deforest07}.  Only the positive flux is measured since the negative flux is difficult to separate from the negative background flux. However,  the magnetic field is divergence free, therefore, the same amount of negative and positive flux is expected to emerge. 

\begin{figure*}   
\epsscale{1.15}
\plotone{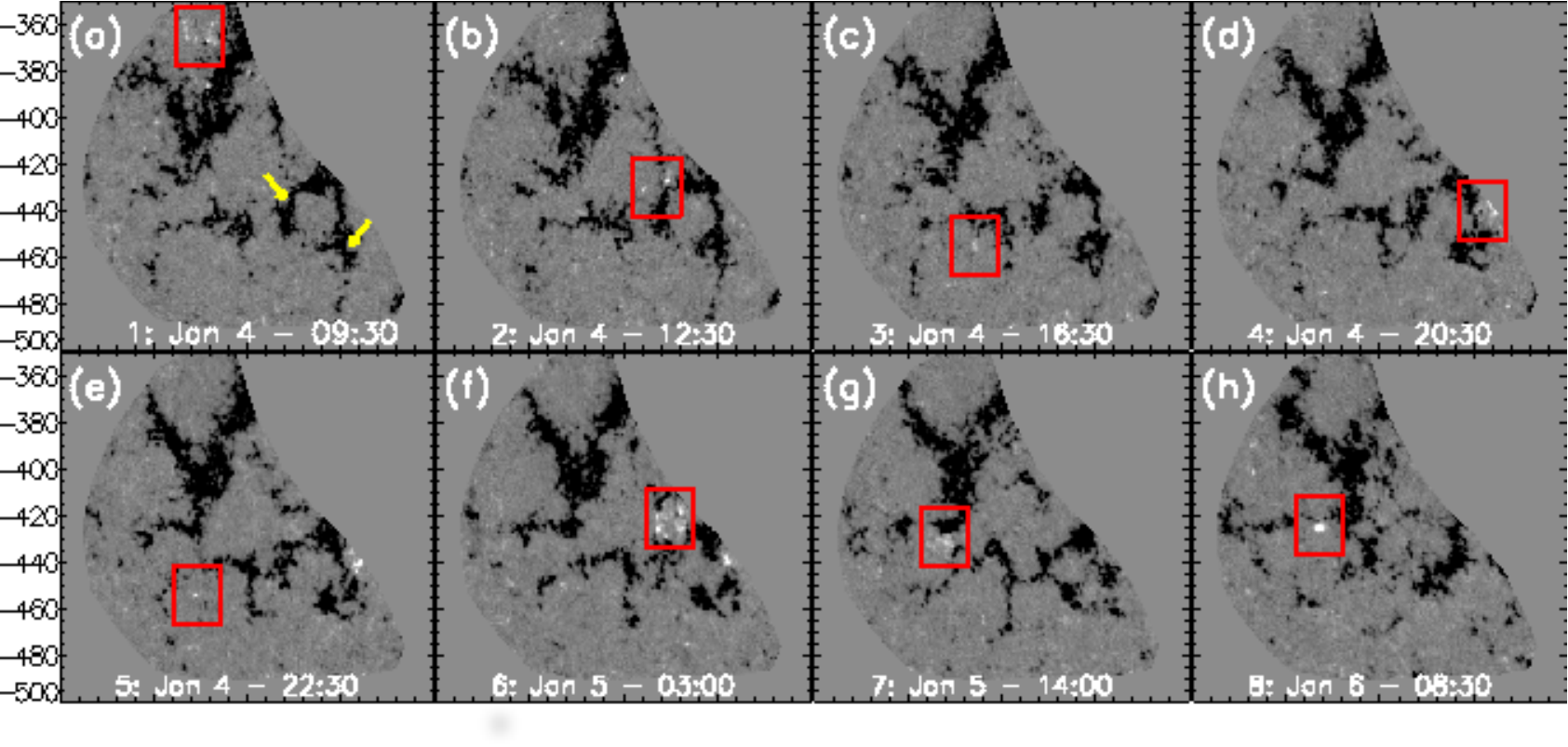}
\caption{SDO/HMI magnetograms of R1 contour showing the locations of flux emergence events. In panel (a) the yellow arrows indicate the branches of the inverse-Y shaped negative field discussed in \sect{results_Supergranular}. 
\label{flux_plot}}
\end{figure*}

Emerging positive magnetic fragments were deemed to be an `event' if the fragments had magnetic flux greater than 10$^{18}$ Mx (singular or collective fragments with a separation of less than 25$\arcsec$), had a duration longer than 30 minutes, and were observed to interact and undergo flux cancellation with the supergranular cell of AR~11389.  Eight events met this criteria.  They are shown in a series of SDO/HMI zoomed magnetograms in \fig{flux_plot}.  Starting/ending times and maximum magnetic flux for each event are given in Table \ref{tab:01}.

\begin{table}[h]
\renewcommand{\arraystretch}{1.5}
  \caption{Positive flux emergence events within R1 from 2012 January 4--6. 
  }
  \vskip4mm
 	\begin{center}
  \label{tab:01}
  \begin{tabular}{cccccc}
  \hline
\\
Event &
Start Time (UT)&
End Time (UT)&
Maximum Positive &

\\
&&&Flux (Mx)
\\
\hline
\\
$1$ &
$4$--Jan $09$:$30$ &
$4$--Jan $13$:$30$ &
$0.9 \times 10^{19}$
 \\
$2$ &
$4$--Jan $10$:$30$ &
$4$--Jan $14$:$30$ &
$0.4 \times 10^{19}$
\\
$3$ &
$4$--Jan $15$:$30$ &
$4$--Jan $18$:$30$ &
$0.2 \times 10^{19}$
 \\
$4$ &
$4$--Jan $18$:$00$ &
$5$--Jan $05$:$30$ &
$1.2 \times 10^{19}$
\\
$5$ &
$4$--Jan $23$:$30$ &
$5$--Jan $11$:$00$ &
$ 0.3 \times 10^{19}$
\\
$6$ &
$4$--Jan $22$:$30$ &
$5$--Jan $03$:$30$ &
$2.0 \times 10^{19}$
 \\
$7$ &
$5$--Jan $12$:$30$ &
$5$--Jan $17$:$30$ &
$0.7 \times 10^{19}$
\\
$8$ &
$6$--Jan $08$:$00$ &
$6$--Jan $10$:$30$ &
$1.2 \times 10^{19}$
 \\
\\
\hline
\end{tabular}
\end{center}
\end{table}

Total negative flux within the contour of R1 at the time of the \emph{Hinode}/EIS raster on the 4th is 2.1$\times$10$^{21}$ Mx.  Over two days, the eight flux emergence events added 6.9$\times$10$^{19}$ Mx of positive flux to R1.  Hence, photospheric flux emergence added $\sim$3\% to the total flux  around the supergranular cell within R1 where we observed an increase in photospheric composition plasma within coronal loops.  This represents $\sim$0.5\% of total flux within AR~11389.  

\begin{figure}   
\epsscale{0.6}
\plotone{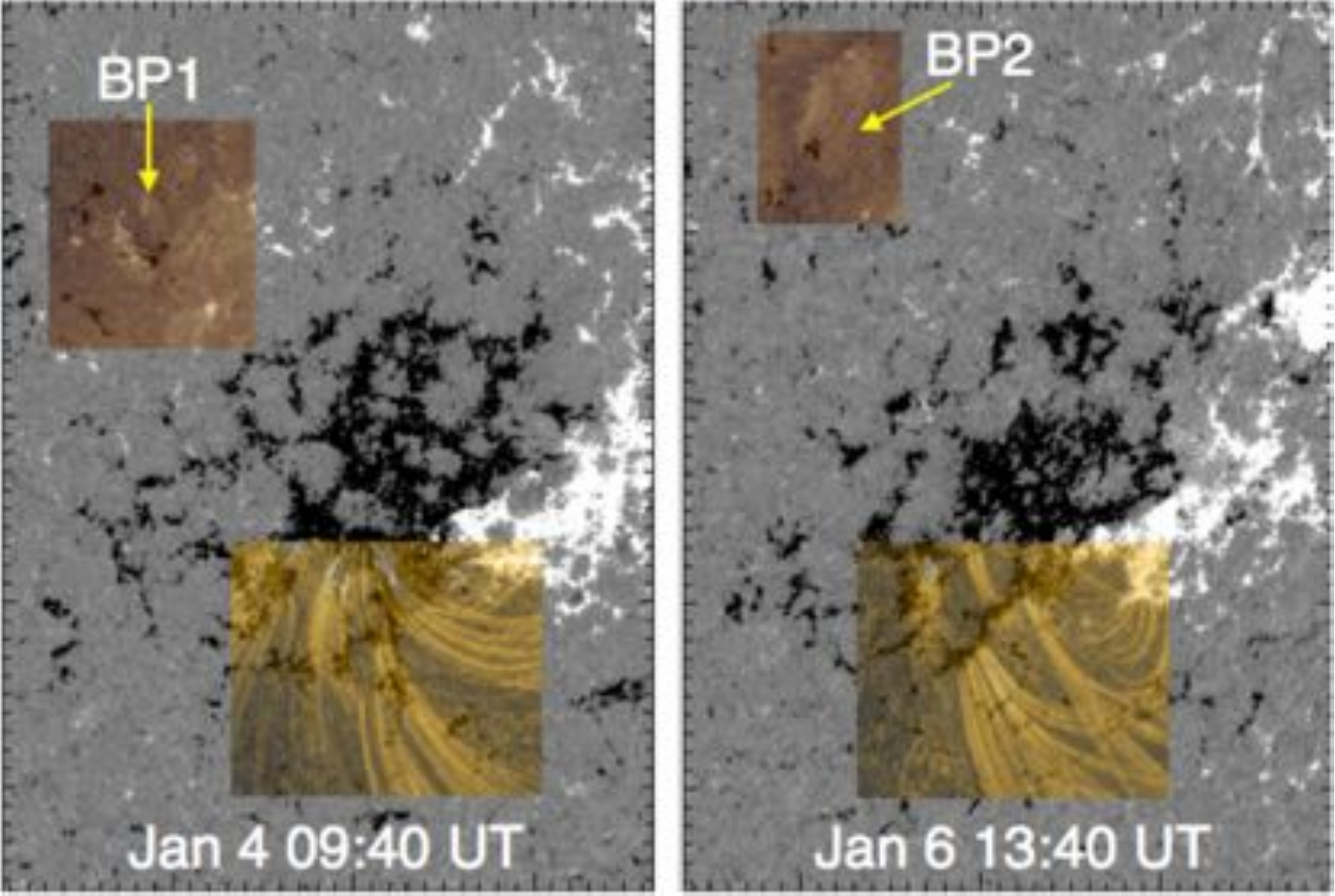}
\caption{SDO/HMI magnetograms corresponding to the EIS FOV for rasters of AR~11389 on January 4 and 6 overplotted with regions of SDO/AIA 171 \AA\ (yellow) and 193 \AA\ (bronze) intensity.  On the 4th, some of the long loops on both sides of the inverted Y-shaped negative field (171 \AA\; see yellow arrows in \fig{flux_plot}) are connected to dispersed positive field of AR~11388 (located outside the field of view to the west as seen in \fig{context}a) while other loops connect with the positive polarity of AR~11389.  Two days later, the right branch of the inverted-Y shaped field no longer exists suggesting that loops originally rooted there were forced to reconnect elsewhere within the AR-CH complex.  Bipoles BP1 and BP2 are located on the northeast of AR~11389 and their coronal loops are shown with overplotted 193 \AA\ intensity.  See both movies.
\label{loops}}
\end{figure}

We suggest that the evolution of the photospheric magnetic field led to reconnection with the overlying coronal field.  This is particularly evident on the western side where the negative magnetic field in the photosphere  forms an inverted-Y on January 4 (see arrows in \fig{flux_plot}a).  However, by the 6th, the right branch of the inverted Y-shaped field is no longer intact (see the evolution in the bottom panels of \fig{flux_plot} and in the upper left panel of SDO$\_$multi$\_$panels.mov).  The left panel of \fig{loops} highlights some long loops rooted on the right side of the inverted Y-shaped negative field on the 4th.  These loops observed in the SDO/AIA 171 \AA\ passband connect outside the \emph{Hinode}/EIS FOV shown, within dispersed positive field of the nearby AR~11388 located towards the west of AR~11389 (\fig{context}a).  Shorter loops rooted nearby connect within AR~11389.  Repeated flux emergence and the subsequent field cancellation has eroded the photospheric magnetic field and forced these coronal loops to reconnect elsewhere by the 6th.  The SDO$\_$multi$\_$panel.mov movie shows the forced reorganization of the coronal loops rooted along the inverted-Y field and the periphery of the supergranular cell in five passbands.  

We conclude that in R1, supergranular cell evolution, characteristic of all decaying ARs, led to the observed decrease in FIP bias.  Small bipoles emerging in and around the supergranular cell reconnected with the pre-existing coronal field.  As a consequence of the reconnection, the photospheric plasma contained within these bipoles mixed with high-FIP bias plasma of the coronal loops.

\subsection{Composition Evolution: Emergence and Spot Decay}
\label{results_Localized}

Changes in FIP bias occur in other localized regions (R2--R4) within AR~11389 and along its north-eastern edge (R5), adjacent to the coronal hole.  The corresponding PDFs of FIP bias pixel$^{-1}$ for R2--R5 are shown in the top panels of \fig{pdfs} and PDFs for subregions within R3 are shown in the bottom panels.  PDFs of all major regions shift to lower levels of FIP bias \emph{i.e.} towards photospheric composition and mean values pixel$^{-1}$ decrease by 0.21--0.31 as plasma evolved from the 4th to the 6th.  

\begin{figure}   
\epsscale{1.0}
\plotone{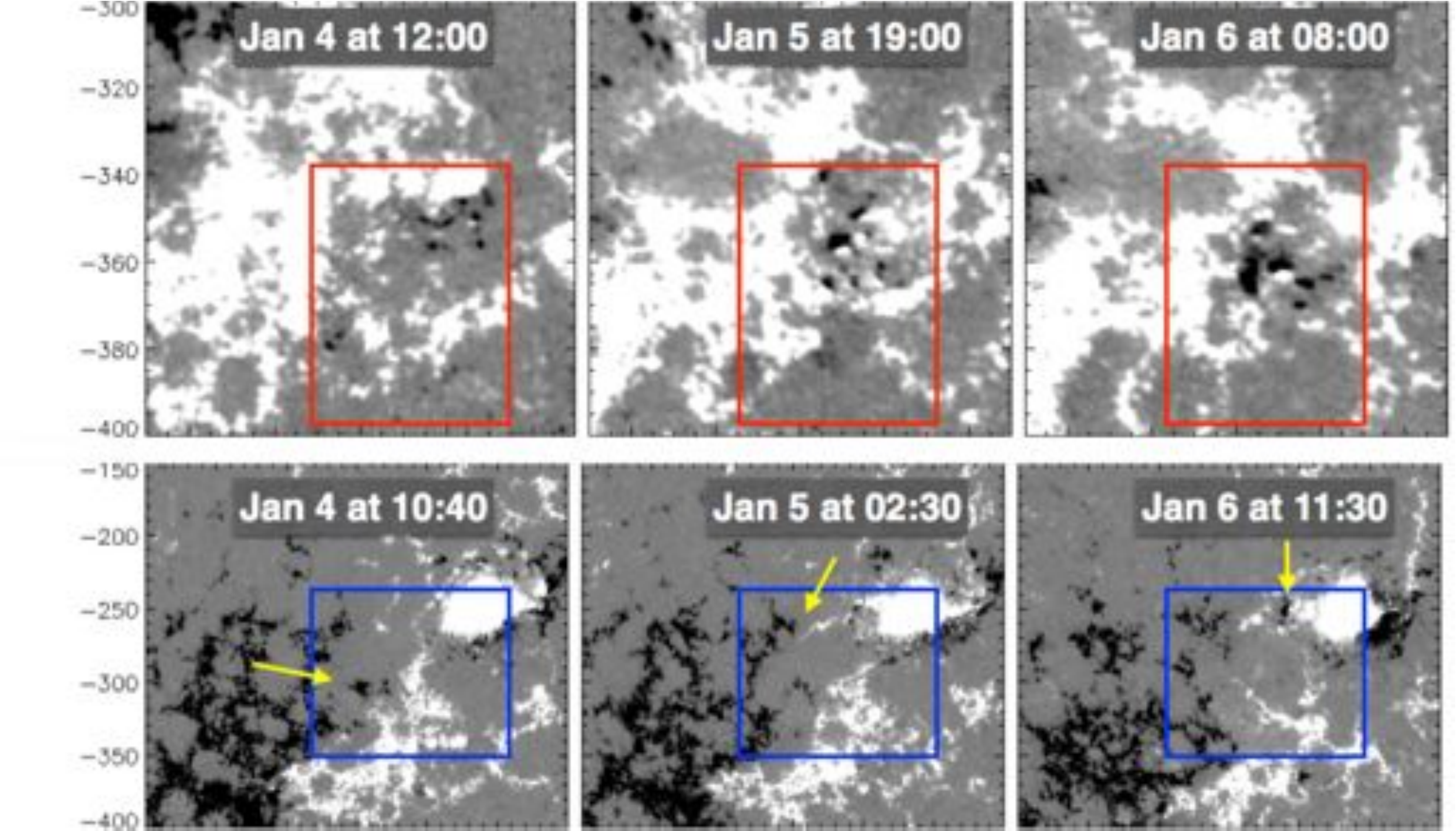}
\caption{SDO/HMI zoomed images showing magnetic field evolution described in \sect{results_Localized}.  Yellow arrows indicate locations of flux cancellation along the fragmented PIL in R4.  The HMI FOV and the sizes of the red/blue boxes do not correspond to the EIS FOV.   \label{reg3}}
\end{figure}

Paths of low-FIP bias in R2 on January 6 appear to trace the new loops that formed as a consequence of reconnection along the supergranular cell boundary described in \sect{results_Supergranular} (\fig{contours}b).  Furthermore, the western-most low-FIP bias patch within R2 is cospatial with the location of persistent and highly concentrated activity in the dispersed positive field at Y$\sim$-350$\arcsec$ to -380$\arcsec$.  Moving magnetic features \citep[MMFs;][]{harvey73} are evident here on the 4th, prior to repeated flux emergence beginning later on the 4th and continuing until the early hours of January 6.   Zoomed SDO/HMI magnetograms showing the evolution of the field in the red boxes are displayed in the top panel of  \fig{reg3}.

Within region R4, small scale magnetic field evolution takes place near the eastern/southeastern side of the large positive spot.  Flux cancellation occurs along the fragmented PIL concentrated in the areas designated by the yellow arrows in the bottom panels in \fig{reg3}.  MMFs feed the process of cancellation as the spot decays throughout the period.  The magnetic field evolution is mostly due to the formation of a supergranular cell (well developed on January~6, \fig{reg3}), although the change observed in FIP bias is not as prominent as in the supergranular cell in the AR's negative field (\sect{results_Supergranular}).

\subsection{Composition Evolution: Eruptions}
\label{results_Eruptions}

FIP bias evolution along the AR's northeastern boundary in R5 is quite stark in the composition maps.  Over two days, the spatial extent of the blue/gray area where FIP bias $\sim$1 has significantly increased, especially towards the south and northwest (\fig{contours}).  Once again, flux emergence has occurred where we have observed an increase in plasma of photospheric composition.  Two small bipoles, BP1 and BP2, have emerged in R5.  \fig{loops} and SDO$\_$multi$\_$panel.mov at 03:00 UT on the 5th show the locations of the bipoles and their overlying/surrounding arcade of loops. 

BP1 is visible in HMI magnetograms a few hours prior to the \textit{Hinode}/EIS observation on January 4.  Relevant frames of SDO$\_$multi$\_$panel.mov have been marked to indicate the sequence of episodes described here. At $\sim$03:00 UT on the 5th, a confined eruption of BP1 begins.  The erupting field subsequently reconnects with the overlying arcade, supplying plasma that is likely to be of photospheric composition and cooler than the plasma contained in the pre-reconnection coronal loops.  This is evident by a darkening in part of the arcade and in the area between BP1 and the northern footpoints of the arcade loops.  The reconnected loops disappear from the SDO/AIA  171 \AA\ images though their `shape' remains.  By 13:00 UT, the system has relaxed and dense coronal loops have reformed and reappear in SDO/AIA 171 \AA\ images as before the confined eruption.  

Similarly, BP2 goes through a confined eruption and and subsequent reconnection with the surrounding arcade field on January 6.  Reconnection is dominantly between closed connectivities (the emerging bipoles and arcade loops) and does not appear to involve the open field of the nearby CH.    

The impact of the confined eruptions of BP1 and BP2 on FIP bias evolution in R5 is noticeable in the composition maps on January 5 and 6 in the middle panels of \fig{all_eis_hmi} (see also red arrows in the first image of FIP$\_$EIS$\_$AIA.mov).  Low-FIP bias plasma extends further south on the eastern side of the AR immediately after the eruption of BP1 on the 5th and expands further south on the 6th after  the eruption of BP2. On the 6th, low-FIP bias plasma also extends along the northern edge of the EIS FOV where reconnected arcade loops are rooted on their northern side.  Clearly, BP1 and BP2 reconnecting with the overlying arcade has provided low FIP-bias material to the arcade loops observed in the composition maps.

\subsection{Composition Evolution: AR Core}
\label{results_Core}

\begin{figure}   
\epsscale{1.0}
\plotone{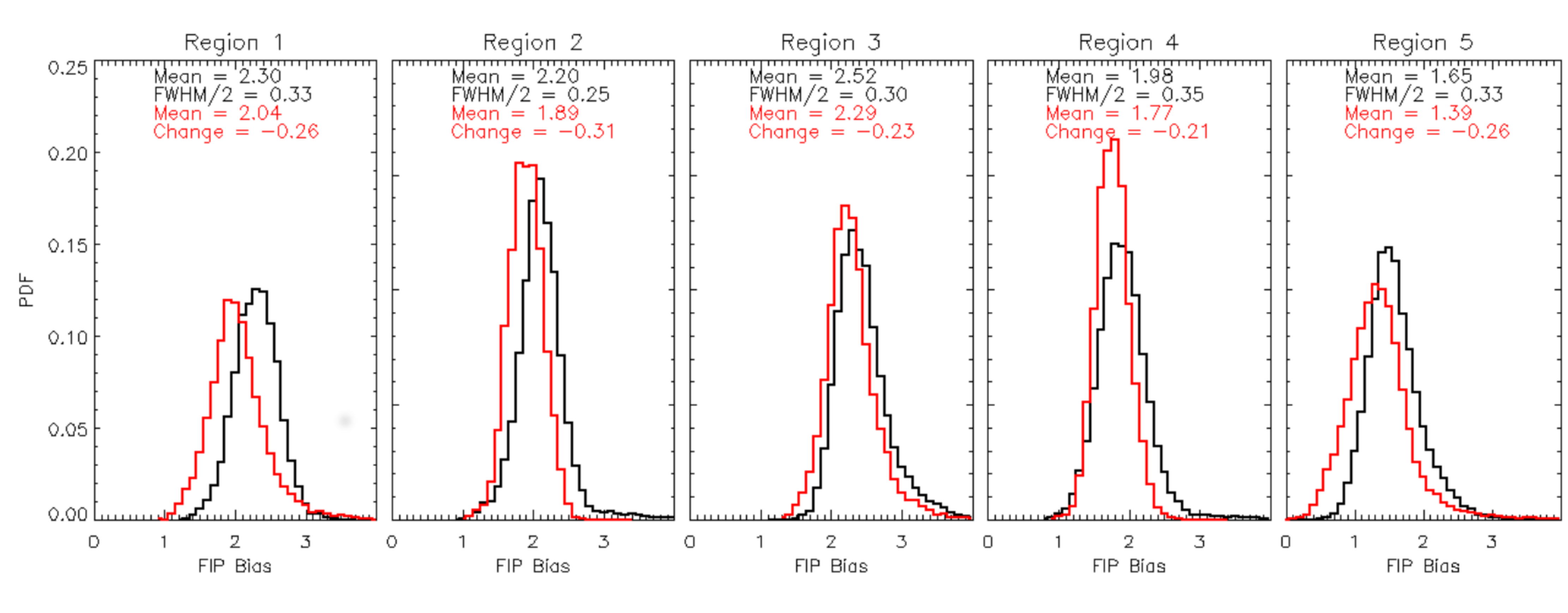}
\plotone{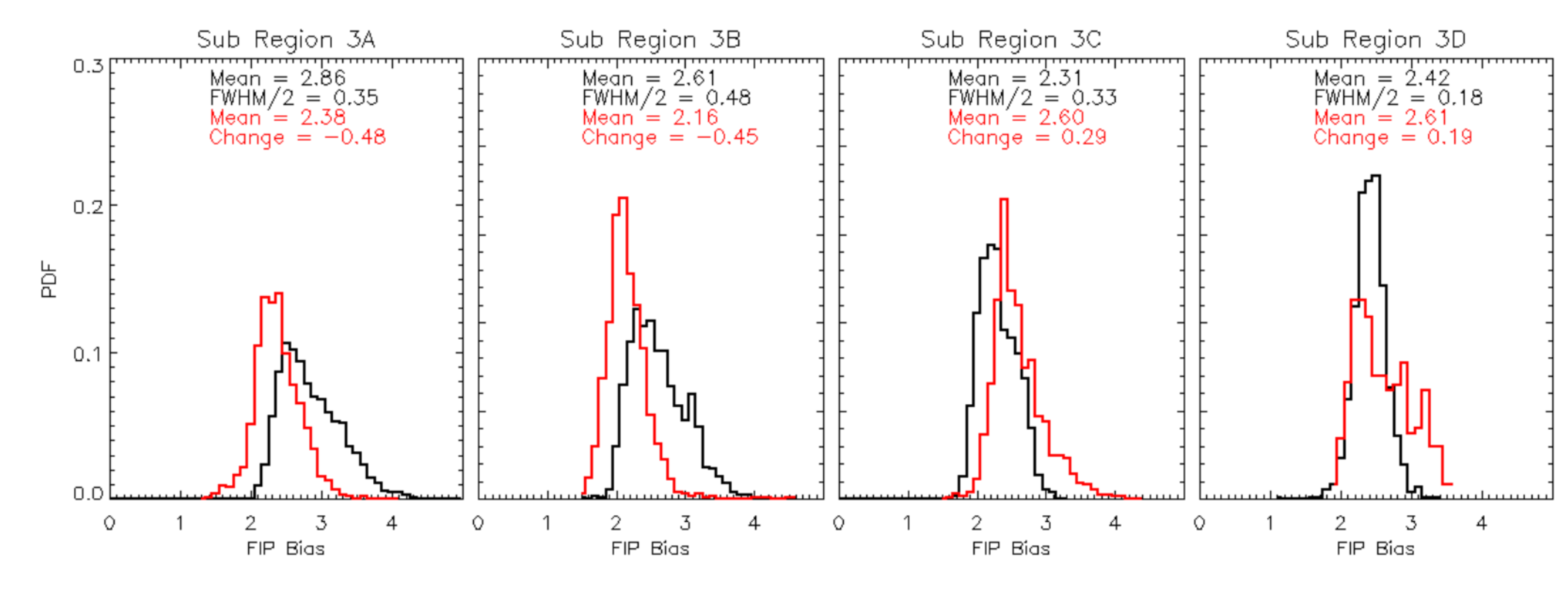}

\caption{Probability distribution functions (PDFs) of FIP bias pixel$^{-1}$ within major regions R1--R5 (top panel) and subregions 3a--3d (bottom panel) for 2012 Jan 4 at 09:40 UT (black) and  2012 Jan 6 at 13:40 UT (red).  (Contours of the regions are shown in \fig{contours}).  \label{pdfs}}
\end{figure}

PDFs of FIP bias in regions R1, R2, R4, and R5 show clear shifts towards photospheric composition after two days (\fig{pdfs}).  However, a more complex case is present within region R3 covering part of the AR core.  Significant localized FIP bias decreases and increases are in fact offsetting each other.  We divided region R3 into four subregions, 3a--3d, for further investigation.  

Contours 3a-3b (red) are located at the footpoints of one set of connecting sheared coronal loops and 3c-3d (blue) are located at the footpoints of a different set of less sheared loops (\emph{c.f.} locations of the subregion contour pairs with the PIL in \fig{contours}).  In the case of 3a-3b, FIP bias substantially decreases with mean values falling 0.48 and 0.45, respectively  (\fig{pdfs}, bottom left panels).  Flux emergence occurs within 3a and within 3b the underlying magnetic flux decreases/disperses as a supergranular cell is evolving in the positive field.  This is consistent with what has been observed in the main regions R4 and R5. 

In contrast, FIP bias increases at the footpoints of the coronal loops connecting 3c-3d (see \fig{pdfs}, bottom right panels).  Magnetic flux density in contours 3c-3d either remains approximately unchanged (3d) or becomes more concentrated (3c).  These regions, 3c-3d, are the only locations within AR~11389 and its surroundings where the FIP-bias enhancement process is not dominated by the faster emergence process.

\section{Discussion}
\label{discussion}
Comparison of a series of \emph{Hinode}/EIS composition maps of AR~11389 from 2012 January 4--6 revealed that plasma composition evolved to lower FIP bias within the AR.  In the major regions, R1, R2, R4, and R5, PDFs of FIP bias pixel$^{-1}$ shifted towards photospheric composition and mean values pixel$^{-1}$ decreased by 0.21--0.31.  Within the AR core (R3) where there was also an observed decrease in FIP bias, substantial localized decreases of 0.48 and 0.45 (subregions 3a--3b) were offset by increases of 0.29 and 0.19 (subregions 3c--3d) in mean values pixel$^{-1}$.  Changes in mean values within one of the main regions, R2, and 2 of the subregions of R3 (3a, 3d) were in excess of half of their full-width-half-maximum (FWHM).  

Overall evolution of plasma composition in AR~11389 differed from that of four ARs in \emph{Skylab} spectroheliograph observations reported by \cite{widing01}. In their sample of newly emerged ARs, FIP bias increased from within hours of flux emergence to reach coronal abundances after two days. The fastest increase in the FIP bias occured during the phase of rapid sunspot growth, from 1--3 days after emergence.  \cite{widing01} concluded that the FIP-effect mechanism is more effective during the emergence phase.  Unlike the young ARs observed by \emph{Skylab}, AR~11389 was a mature AR in its decay phase.  As expected in decaying ARs, the evolution of underlying magnetic field was dominated by the formation of supergranular cells and small-scale emergence/cancellation/MMFs which we observed in major regions R1, R2, R4, and R5, and subregion 3a--3b, the same regions where FIP bias decreased over two days. 

In contrast, FIP bias increased in contours 3c--3d where the magnetic field evolved differently compared to the other areas in the AR.  Magnetic flux density remained concentrated at each loop footpoint rather than decreasing by dispersion and cancellation.  The fact that both FIP bias and the magnetic field evolved in opposite directions in these locations compared to regions 3a--3b, also in the AR core, provides a reasonable degree of confidence that systematic errors are not significant.  However, as was stated in \cite{baker13}, Si, with FIP of 8.2 eV, and S, with FIP of 10.4 eV, are close to the boundary between high and low FIP elements on the Sun at 10 eV.  \cite{lanzafame_etal2002} showed that S behaves like a high-FIP element in ARs but it may over-fractionate in quiet Sun regions \citep{lanzafame05,brooks_etal2009}.  Thus the behavior of S may be sensitive to the region on the Sun being investigated.  For a discussion of caveats associated with the S {\sc x} 264.223 \AA\ -- Si {\sc x} 258.375 \AA\ and a test of the validity of the method please see \cite{baker13} and \cite{brooks15}. 

\cite{sheeley95,widing97,young97,widing01} found photospheric plasma composition in newly emerged loops, therefore, flux emergence provides a potential reservoir of low-FIP bias plasma to mix with the high-FIP bias plasma contained within AR coronal loops.  The key questions are how does this material make its way into coronal loops of the AR and on what time scales does the plasma mixing occur?  As was demonstrated in \cite{baker13}, reconnection can provide the pathways for mixing of magnetic field containing plasma of different compositions.  
 
When magnetic flux emerges it displaces the pre-existing field in the AR, a current sheet generally forms at the boundary between the different magnetic domains, and reconnection takes place forming new coronal magnetic connections \citep[e.g.][ and references therein]{guo13,tarr14}.  This implies that small scale flux emergence within supergranular cells and flux cancellation along their boundaries will involve reconnection.  The effect of MMFs is similar to that of small scale flux emergence episodes.  Both involve the interaction between new, small-scale loops and large, pre-existing AR loops.  In this way, photospheric plasma from the flux emergence episodes can be injected into the pre-existing loops of the AR and surrounding arcade containing high-FIP bias plasma.  In AR~11389, evolution of supergranular cells, especially in the dispersed negative field, leads to the reorganization of the overlying field from January 4--6.  SDO/AIA and SDO/HMI observations show that dispersing/disappearing field forces loops to reconnect within the AR, the surrounding arcade, and network.  Then, low-FIP bias plasma fills the reconnected arcade loops.

The plasma composition of AR~11389 is likely to be determined by the relative time scales of plasma mixing resulting from small-scale flux emergence and FIP bias enrichment processes.   \cite{widing01} observationally determined that such enrichment in newly emerged ARs occurs over days.  This is consistent with the time scales of the ponderomotive force FIP effect model of \cite{laming04,laming09,laming12}.  The ponderomotive force creates local abundance variation in relatively short time scales ($\sim$10$^{3}$ s) at chromospheric heights.  As a result, observed coronal abundance anomalies require either transport or diffusion processes in order to supply low FIP ions to the corona.  Typical timescales for these processes are of order of days \citep{laming14}.   

Plasma mixing time-scales due to small-scale flux emergence and subsequent cancellation episodes within the evolving magnetic field of supergranular cells are likely to be short compared to those of FIP bias enrichment and transport processes.  These mixed polarity features are highly dynamic and evolve on a time scale of few hours (Table \ref{tab:01}).  Within the supergranular cell of the AR's negative field, we observed numerous episodes with lifetimes of 1--12 h and maximum flux emergence per episode in the range $\sim$0.2--2$\times$10$^{19}$ Mx.  Despite their small spatial scales and short lifetimes, they contribute a significant fraction to the total solar magnetic flux \citep[see][ and references therein]{thornton11}.  For BP1 and BP2, the process is similar but more efficient as the eruptions forced more reconnection for longer time periods. 

\section{Conclusions}
\label{Conclusion}

In this paper, we provide clear evidence of FIP bias evolution in a decaying AR near a CH from 2012 January 4--6 using SDO/AIA, SDO/HMI, and \emph{Hinode}/EIS observations.  A sequence of large FOV, high spatial resolution S {\sc x}--Si {\sc x} EIS composition maps shows a global increase in plasma of photospheric composition in AR~11389.  We analyzed in detail the locations of significant changes in FIP bias and found that small-scale evolution of the underlying magnetic field is closely linked to the evolution of plasma composition observed in the corona.  

For all regions and subregions where FIP bias decreased, we identified episodes of flux emergence/cancellation and moving magnetic features (MMFs). We quantified the amount of flux added to the region containing part of the well-formed supergranular cell in the ARÕs negative field. What appeared to be small-scale, intermittent flux emergence episodes in fact added $\sim$3$\%$ of new flux to region R1 and $\sim$0.5$\%$ to the entire AR over two days. This is small compared to the large-scale flux emergence of 13$\%$, which was added over two days to decaying AR~11112 \citep{tarr14}.

It is widely accepted that active regions begin decaying almost as they emerge into the photosphere.  The decay process spreads the concentrated magnetic flux over an ever-increasing area due to supergranular buffeting.  Supergranular flows advect the magnetic field to cell edges creating `holes' of relatively free-field regions.  Most of the flux disappears through small-scale emergence and subsequent cancellation \citep{lvdg14}.  MMFs also contribute to the removal of magnetic field as opposite polarities cancel \citep{harvey73,hagenaar05}. 

We observed the effect on plasma composition of small-scale flux emergence within supergranular cells and flux cancellation along their cell boundaries in AR~11389.  High-FIP bias is conserved/amplified only in localized areas of high magnetic flux density \emph{i.e.} in the AR core.  In all other areas within the AR, FIP bias decreased.  During the normal course of the AR decay process, photospheric composition plasma is supplied to coronal loops containing high-FIP bias plasma via reconnection of new emerging loops and the pre-existing field on time scales (hours) that are shorter than those of enrichment processes (days).  This is in contrast to repetitive heating episodes in loops where high-FIP bias plasma concentrated at loop footpoints \citep{baker13} is transported into the corona by chromospheric evaporation.  Such a process is occurring in young ARs where strong magnetic field provides substantial magnetic energy released in many episodes.  Heating, which is a function of magnetic flux density, is strongly decreasing with the dispersing fields in decaying ARs \citep{lvdg03,demoulin03}.  

\textit{Hinode}/EIS has provided a new opportunity to re-examine the evolution of plasma composition in ARs at the highest temporal and spatial resolutions.  We have extended the work of \cite{widing01} based on \emph{Skylab} observations of very young ARs to the later, decaying phase of a much older AR.  In an aging AR, FIP bias is modulated by small-scale flux emergence.  This result has possible implications for the FIP bias levels observed in the solar wind.  Slow wind FIP bias measured in situ is a steady 3--4 \citep[e.g.][]{geiss95,zurbuchen06}.  Values in excess of or below 3--4 are rare, suggesting that very high-FIP bias measured in ARs ($>$5) does not find a pathway into the slow SW.   Then, only mature and not too dispersed ARs are expected to contribute to the slow SW. 


\acknowledgements{\textit{Hinode} is a Japanese mission developed and launched by ISAS/JAXA, collaborating with NAOJ as a domestic partner, NASA and STFC (UK) as international partners. Scientific operation of \textit{Hinode} is by the \textit{Hinode} science team organized at ISAS/JAXA. This team mainly consists of scientists from institutes in the partner countries. Support  for the post-launch operation is provided by JAXA and NAOJ (Japan), STFC (U.K.), NASA, ESA, and NSC (Norway).  LvDG and DML acknowledge the European Community FP7/2007-2013 programme through the eHEROES Network (EU FP7 Space Science Project No.284461).  LvDG acknowledges the Hungarian government for grant OTKA K 109276.  The work of DHB was performed under contract with the Naval Research Laboratory and was funded by the NASA \textit{Hinode} program.   DB thanks STFC for support under Consolidated Grant ST/H00260/1.  We thank the referee for his/her constructive comments, which have improved the manuscript.
}

\bibliographystyle{apj}
\bibliography{test} 



\end{document}